\documentclass[preprint,review,12pt]{elsarticle}

\usepackage{amssymb}
\usepackage{amsthm}
\usepackage{amsmath}
\usepackage{tablefootnote}
\usepackage{xcolor}

\usepackage{lineno}
\usepackage{makecell}

\journal{arXiv}

\begin{document}

\begin{frontmatter}



\title{Exploring the Correlation Between Ultrasound Speed and the State of Health
of LiFePO$_4$ Prismatic Cells}


\author[inst1]{Shengyuan Zhang}

\affiliation[inst1]{organization={School of Mechanical and Aerospace Engineering, Nanyang Technological University},
            addressline={50 Nanyang Avenue}, 
            country={Singapore 639798}}

\author[inst2]{Peng Zuo}

\affiliation[inst2]{organization={Advanced Remanufacturing and Technology Centre, Agency for Science, Technology and Research}, addressline={3 Cleantech Loop}, country={Singapore 637143}}

\author[inst3]{Xuesong Yin}

\affiliation[inst3]{organization={Institute of Materials Research and Engineering, Agency for Science, Technology and Research}, addressline={2 Fusionopolis Way}, country={Singapore 138634}}

\author[inst1]{Zheng Fan\corref{cor1}}

\cortext[cor1]{Corresponding author}
\ead{zfan@ntu.edu.sg}

\begin{abstract}
Electric vehicles (EVs) have become a popular mode of transportation, with their performance depending on the ageing of the Li-ion batteries used to power them. However, it can be challenging and time-consuming to determine the capacity retention of a battery in service. A rapid and reliable testing method for state of health (SoH) determination is desired. Ultrasonic testing techniques are promising due to their efficient, portable, and non-destructive features. In this study, we demonstrate that ultrasonic speed decreases with the degradation of the capacity of an LFP prismatic cell. We explain this correlation through numerical simulation, which describes wave propagation in porous media. We propose that the reduction of binder stiffness can be a primary cause of the change in ultrasonic speed during battery ageing. This work brings new insights into ultrasonic SoH estimation techniques.
\end{abstract}

\begin{graphicalabstract}
\includegraphics{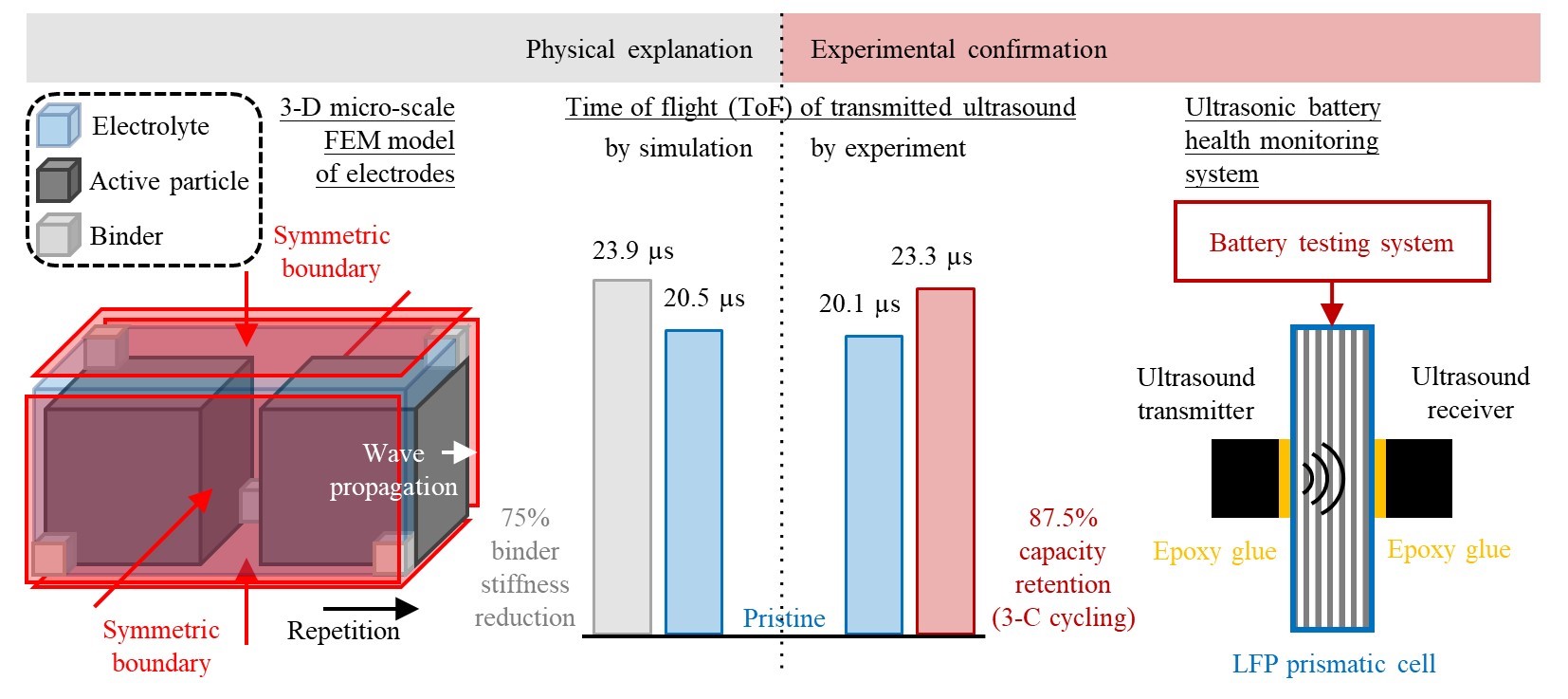}
\end{graphicalabstract}

\begin{keyword}
State of health \sep ultrasonic testing \sep Biot theory \sep finite element method \sep Li-ion battery \sep electric vehicle
\end{keyword}

\end{frontmatter}


\section{Introduction}
\label{sec:intro}
Nowadays, there is explosive growth in the electric vehicle (EV) market due to increased public awareness of environmental protection and improvements in EV range and safety. To achieve a range comparable to traditional vehicles, an EV is usually equipped with hundreds/thousands of Li-ion batteries. However, these batteries will experience ageing in terms of capacity reduction and suffer a high risk of failure. To ensure the range and safety of EVs, the batteries are usually replaced when their capacity retention decreases to 70-80\%.

The capacity of a battery can be measured by applying a full charge-discharge process \cite{lipu2018review}. However, this process can be time-consuming \cite{berecibar2016critical} and the batteries in service usually do not experience full charges or discharges.

In addition to the tools based on electrochemical \cite{berecibar2016state,zhu2022data} and pressure \cite{huang2022onboard} measurements, ultrasonic testing techniques are promising due to their sensitive, portable, and non-destructive features. Pioneering studies on the ultrasonic state of charge (SOC) estimation \cite{hsieh2015electrochemical, hodson2021initial, sun2022ultrasonic, meng2022robust, ke2022potential,huang2023precise} have been reported. Besides, Huang \textit{et al.} \cite{huang2022quantitative} characterized the layered structure within batteries with ultrasonic resonance; Zhang \textit{et al.} \cite{zhang2022ultrasonic} demonstrated the feasibility of using ultrasonic guided waves to monitor dendrite formation at electrode-electrolyte interface in aqueous zin ion batteries; Appleberry \textit{et al.} \cite{appleberry2022avoiding} used ultrasound to detect overcharge and thermal runaway of Li-ion batteries.

For SoH estimation, Davies \textit{et al.} \cite{davies2017state} experimentally confirmed a positive correlation between ultrasound speed/amplitude and SoH of LCO pouch cells within 60 cycles, Ladpli \textit{et al.} \cite{ladpli2018estimating} observed a negative correlation between velocity/amplitude of ultrasonic guided waves and SoH of NMC pouch cells before the capacity retention decreased to 96\%, and Wu \textit{et al.} \cite{wu2019ultrasonic} found that the ultrasound time-of-flight (ToF)/amplitude increases along with the cycle number of LCO pouch cells before the SoH degraded to 98\%. Data-driven methods that modelled the correlations have been adopted as a commercial SoH estimation technique \cite{bhanu2018systems, steingart2021determination, ladpli2022battery}. However, the comprehensive investigation of the underlying physics behind the correlation between SoH and ultrasonic features is still limited. While Deng \textit{et al.} \cite{deng2020ultrasonic} has established that the correlation is primarily linked to unwetting in pouch cells, the inherent complexity of batteries often gives rise to multiple fading mechanisms and the sensitivity to ultrasound can vary depending on the specific case. For instance, in prismatic cells, there is typically an abundance of electrolytes, making the phenomenon of 'unwetting' less significant in this particular case.

Without a basic understanding of the underlying physics, it is unclear in which cases the correlation holds, such as the disappearance of the correlation between ultrasound amplitude and cycle number after 60 cycles \cite{wu2019ultrasonic}. Moreover, as the correlation varies with cells and the modes of ultrasonic waves \cite{davies2017state,ladpli2018estimating,wu2019ultrasonic}, the transferability of a data-driven model would be challenging, while a physics-based method can be easily adapted for various cases. Therefore, it is necessary to figure out which ageing mechanisms that accompany capacity fading should account for the evolution of ultrasonic features.

\begin{figure}
    \centering
    \includegraphics{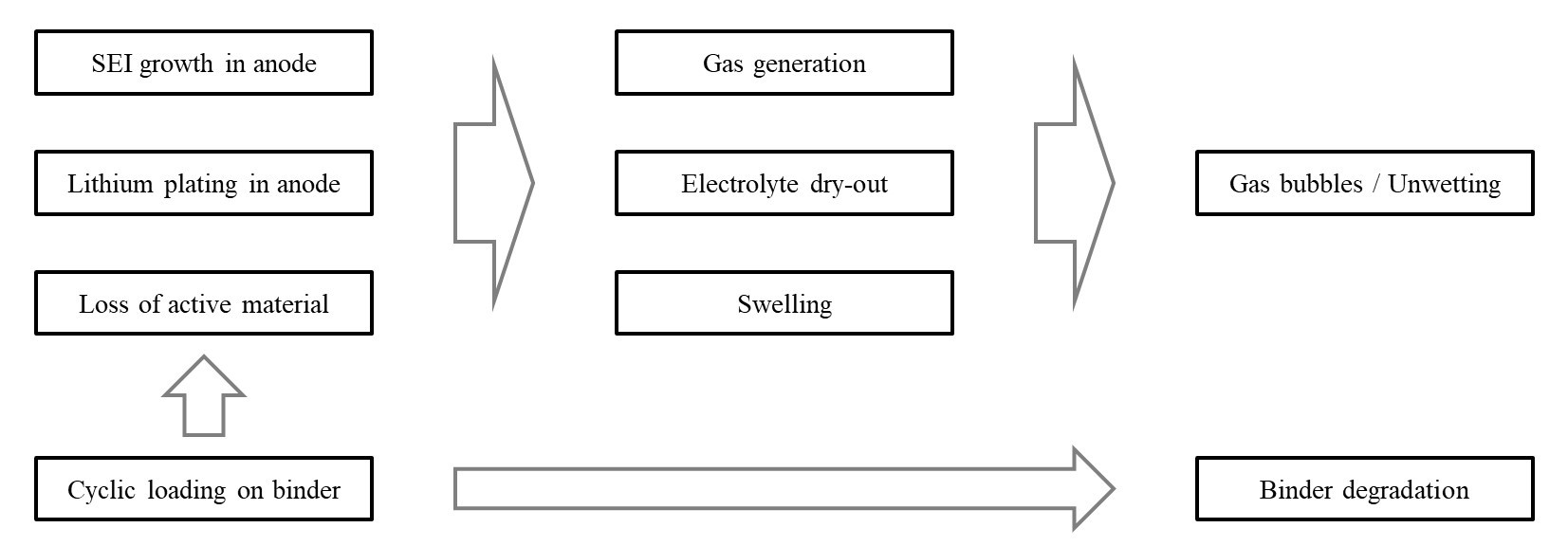}
    \caption{Summary of ageing mechanisms related to structural change and mechanical property evolution.}
    \label{fig:agemech}
\end{figure}

The ageing mechanisms that may introduce geometry and mechanical property changes, and affect ultrasound propagation subsequently, should be considered. As summarized in Figure \ref{fig:agemech}, firstly, capacity fading directly reveals the loss of active lithium. The active lithium can be consumed in three ways: solid-electrolyte interface (SEI) growth and lithium plating in the anode \cite{purewal2014degradation,li2001studies,zhang2011cycling,pinson2012theory,kim2013capacity,zheng2015correlation,rao2019investigation,liang2021numerical}, and loss of active material (LAM) caused by the breakage of the conductive binder. As either graphite particles or cathode active material would expand/shrink during lithium intercalation/deintercalation \cite{qi2010threefold,park2012principles,purewal2014degradation,clerici2021experimental} and the relative amount of intercalated lithium affects their elastic properties \cite{qi2010threefold,maxisch2006elastic}, the loss of active lithium causes variations in particle volumes and elastic properties. Secondly, due to the expansion/shrinkage of active particles during lithium intercalation/deintercalation, the binder that connects the particles suffers cyclic loading \cite{clerici2021experimental,dai2017state}. Demirocak \textit{et al.} \cite{demirocak2015probing} conducted nanoindentation on the cathodes of LFP prismatic cells and found that the cathode solid frame’s modulus degradation due to cycling can be up to 76.1\% under 90\% capacity retention and 2 C-rate. They speculated that the modulus degradation should be attributed to the crystallinity reduction of the binder. This speculation is reasonable because plastics would experience stiffness reduction under cyclic loading due to the breakage of cross-links between molecules on the nano-scale \cite{hwang1986fatigue,gamstedt1999fatigue,thwe2003durability,thornton2007fatigue,lambers2013microdamage}. Moreover, Jiang et al. \cite{jiang2020machine} revealed the detachment between active particles and carbon/binder matrix by experiment, which would not only cause LAM but also the stiffness reduction of the electrodes. Another important consideration is the coupling of SEI growth and lithium plating with gas generation \cite{michalak2015gas,li2023inhibiting} and electrolyte dry-out \cite{deng2020ultrasonic}. These fading mechanisms can lead to the formation of dilute gas bubbles and the emergence of large gas regions, commonly referred to as unwetting pores. Both forms may affect the wave velocities \cite{corapcioglu1991wave,smeulders1997wave,deng2020ultrasonic}. To quantify their contribution to the evolution of ultrasonic features, a model that describes the ultrasound propagation is required. Such a model can not only explain the correlation but also be useful for SoH prediction.

The main challenge of constructing a physics-based model is to simulate wave propagation in electrodes and separators, which are liquid-saturated porous mediums. While an analytical solution for wave velocity in porous media can be derived using Biot theory \cite{carcione2007wave}, its application to electrodes poses significant challenges. Biot theory is primarily designed for analyzing porous media with isotropic and homogeneous solid frames. However, the frame of electrodes exhibits a diverse composition, comprising various materials, including active components and binders. As an alternative approach, we intend to employ 3-D FEM to model the microstructure of the electrodes and simulate the fluid-solid interaction at the microscopic level. This method will enable us to accurately model wave propagation within the electrodes. Similar modelling strategies were adopted in ultrasonic diagnosis for cancellous bone \cite{bossy2011numerical,vafaeian2014finite} and studying seismic wave propagation \cite{van2002finite}.

This paper is organized as the following: Experimental confirmation of the correlation between ultrasound speed and SoH of a commercial LFP prismatic cell for EVs is described in Section \ref{sec:exp}. The physical model construction and the evaluation of the ageing mechanisms to explain the correlation are described in Section \ref{sec:explain}. Section \ref{sec:con} includes the concluding remarks.

\section{Experimental confirmation}
\label{sec:exp}

\begin{figure}
    \centering
    \includegraphics{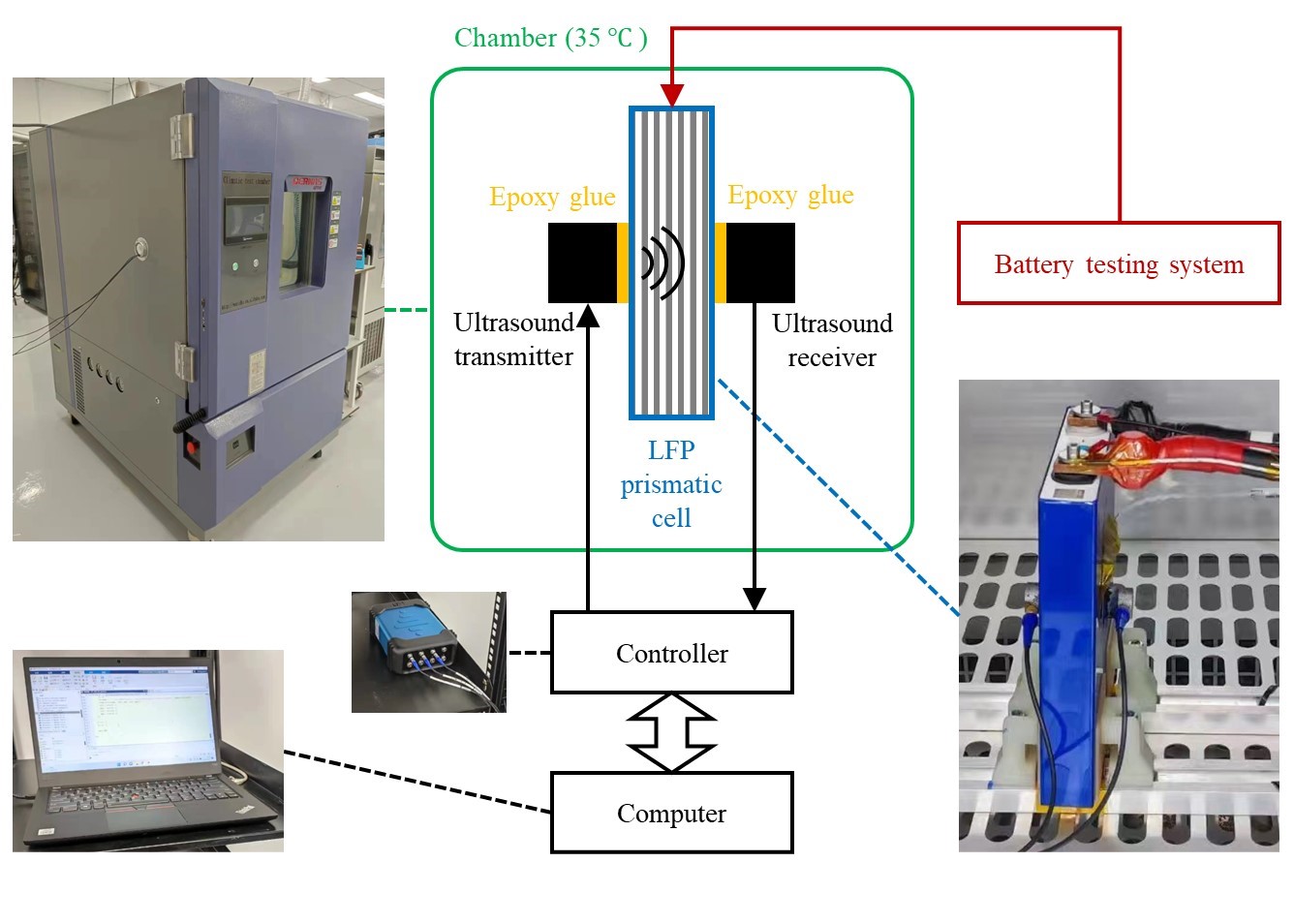}
    \caption{The experimental setup of the ultrasonic battery monitoring system.}
    \label{fig:exp}
\end{figure}

LFP prismatic cells are widely used for EVs. We performed an ageing test on an LFP prismatic cell with a rated capacity of 50 Ah (EVE LF50K) and equipped the cell with an ultrasonic monitoring system concurrently to investigate the correlation between the SoH and ultrasound features. The whole setup is as shown in Figure \ref{fig:exp}.

\begin{figure}
    \centering
    \includegraphics{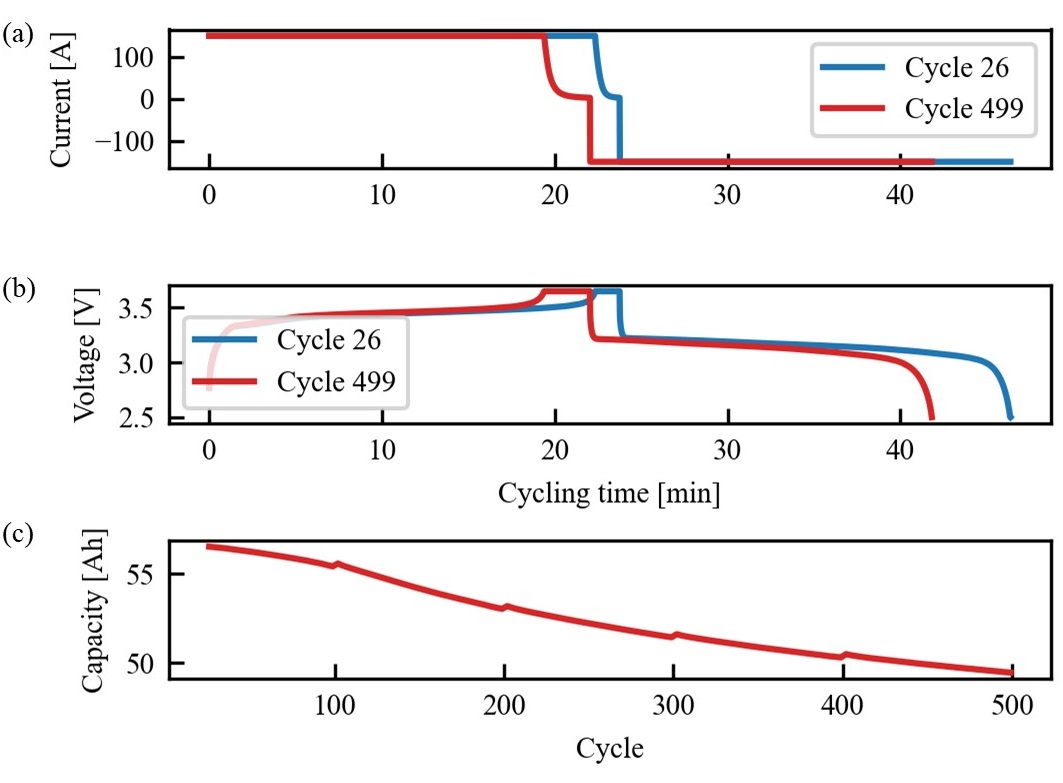}
    \caption{(a) Current and (b) voltage in each charge-discharge cycle; (c) Capacity retention over cycles.}
    \label{fig:aging}
\end{figure}

The cell was connected to a battery testing system (BTS) to charge and discharge at a 3 C-rate, i.e., a rate under which 1/3 h is needed for a full charge/discharge. As shown in Figure \ref{fig:aging}(a) and (b), at each charge-discharge cycle, the cell experienced constant current (CC) charge at 150 A until 3.65 V, constant voltage (CV) charge until 2.5 A, and CC discharge at 150 A until 2.5 V. The capacity fading of the battery due to cycling is shown in Figure \ref{fig:aging}(c). There was a resting stage every 100 cycles, after which there is a capacity regeneration due to the lithium redistribution in different forms \cite{olivares2012particle,qin2016rest}. After 500 cycles, the battery suffers a roughly 10\% capacity loss. 

\begin{figure}
    \centering
    \includegraphics{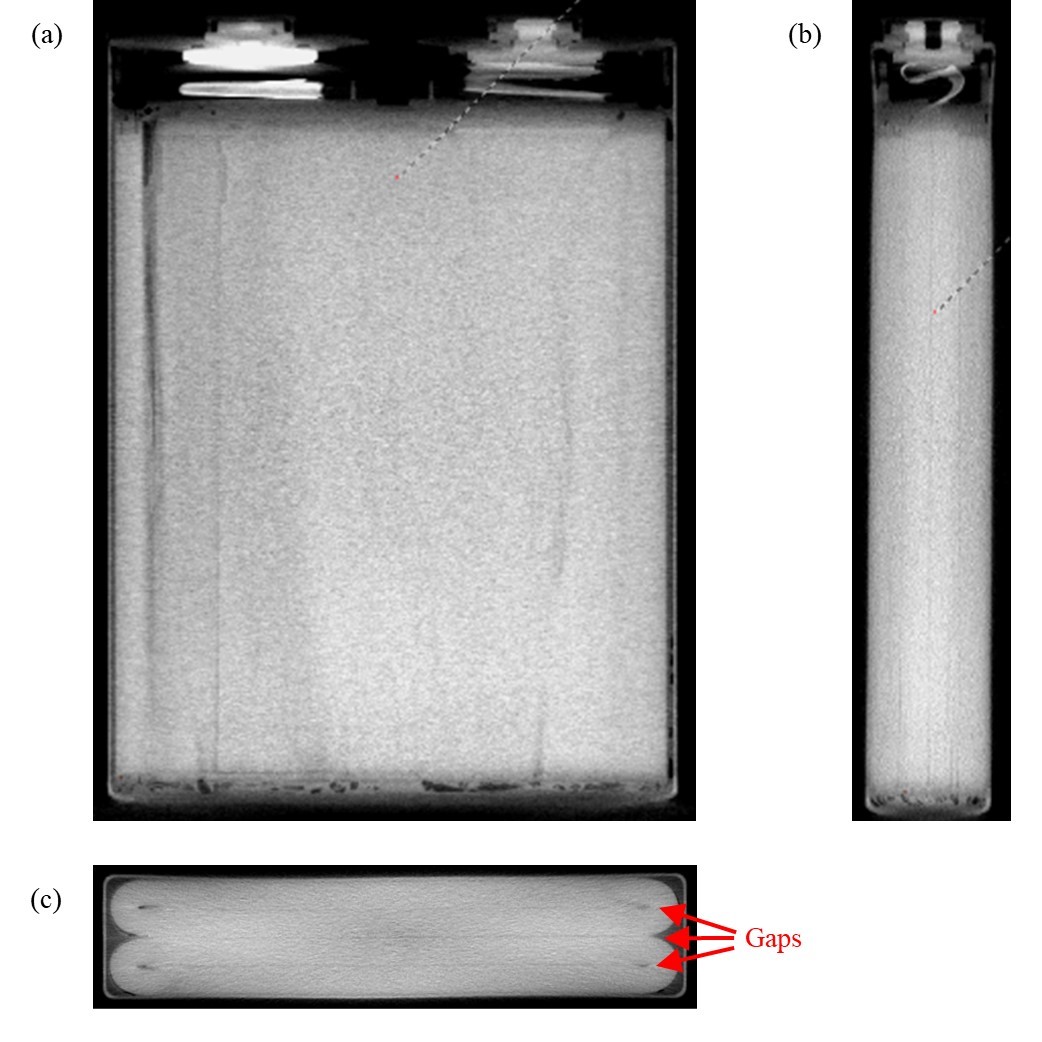}
    \caption{(a) Front view, (b) side view, and (c) top view of the prismatic cell by X-ray computed tomography (XCT) with XTH450 (450 kV, resolution 100 $\mu$m).}
    \label{fig:xct}
\end{figure}

\begin{figure}
    \centering
    \includegraphics{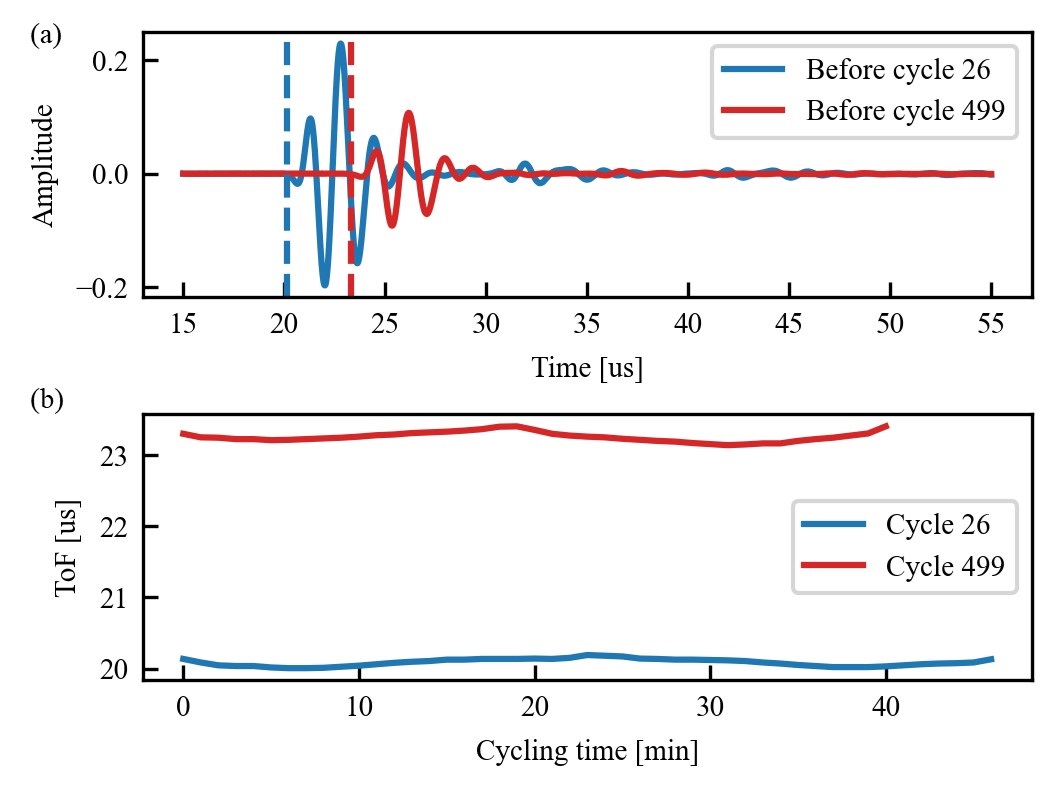}
    \caption{(a) Two sample signals of transmitted ultrasound, the dashed vertical line indicates the time of flight (ToF) of the ultrasound; (b) ToF evolution within a charge-discharge cycle.}
    \label{fig:result}
\end{figure}

Two ultrasound probes (Olympus V106-RM) were bonded on the front and back sides of the battery to generate ultrasonic waves and receive the transmitted ultrasonic waves, respectively. The excitation was a broadband pulse with a central frequency of 2.25 MHz. As illustrated by the X-ray CT images in Figure \ref{fig:xct}, the cell contains two rolls of stacked electrode/separator layers, which are immersed in electrolytes. The waves propagated along the stacking direction of the internal layers. The time of flight (ToF) of the transmitted waves was extracted from the time-domain signals, as shown in Figure \ref{fig:result}(a). After nearly 500 cycles, the ToF of the transmitted waves increased by around 3 $\mu$s. It is worth noting that the central frequency of the received waves was around 600 kHz, which is lower than the excitation frequency. It is expected, as waves with higher frequencies had lower transmission through the cell. Moreover, in addition to the main pulse, the subsequent minor pulses corresponded to the multiple reflections caused by the shells and gaps between the rolls, as shown in Figure \ref{fig:xct}(c).

From Figure \ref{fig:result}(b), we observe that the ToF variation is small in each charge-discharge cycle and the pattern of evolution over time is different between cycles. It implies that ToF is insensitive to SOC and susceptible to other factors.

\begin{figure}
    \centering
    \includegraphics{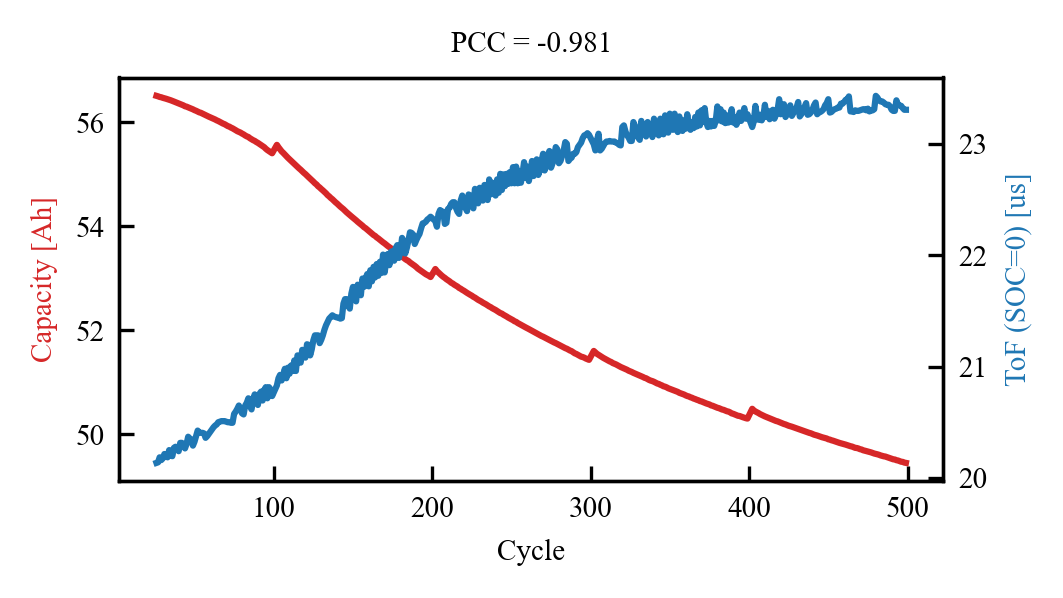}
    \caption{The ToF corresponding to the fully discharged state and capacity retention over cycles.}
    \label{fig:result_}
\end{figure}

On the other hand, as shown in Figure \ref{fig:result_}, ToF has a strong correlation with SoH. Although there are small perturbations, the ToF increased monotonically along the cycle number, which is sound for accurate SoH prediction with ToF. Additionally, the ToF was found to be insensitive to capacity regeneration after each resting stage. To quantify the correlation, a metric that measures the linear correlation between two variables, Pearson correlation coefficient (PCC) \cite{cohen2009pearson}, is adopted. A PCC value close to -1 represents a strong negative linear correlation. The PCC between the capacity retention and the ToF over the 500 cycles is -0.981, which indicates a strong negative linear correlation.

It is also worth noting that, although the chamber kept an ambient temperature of 35 $^\circ$C (see Figure \ref{fig:exp}), during the continuous 3 C-rate cycling, the surface temperature of the cell elevated and equilibrated in the range of 44 to 46.5 $^\circ$C, according to the measurement by a bonded thermocouple. At the fully discharged states, the surface temperature remained constant, which was 46 $^\circ$C. Therefore, the temperature effect on the ToF was excluded.

In summary, from Figure \ref{fig:result_}, the capacity suffers a 12.4\% decrease from 56.5 Ah at the 26th cycle to 49.5 Ah at the 500th cycle under 3 C-rate, and the ToF increases by 15.9\% from 20.1 $\mu$s to 23.3 $\mu$s. Similar results were also found in another cell, which is described in \ref{sec:exp2:appendix}. The strong correlation between ToF and capacity fading indicates the feasibility of using ultrasound for battery health monitoring. We can model the correlation using either data-driven methods or physical models for SoH prediction in practice.

\section{Physical explanation}
\label{sec:explain}

\subsection{Wave propagation model of the prismatic cell}

To quantitatively explain the correlation between ToF and SoH, it is necessary to build up a mathematical model for the cell to calculate the transmission ToF of ultrasound. The model built here is for a pristine cell in a fully-discharged state.

The internal structure of the prismatic cell is stacked layers. Among them, the current collectors (Cu and Al) are simple linear elastic solids, while the separator is fluid-saturated porous material whose solid phase is polypropylene (PP) and fluid phase is the electrolyte. The anode and cathode are porous materials whose solid phase consists of several materials, as listed in Table \ref{tab:binder}, and the electrolyte fluid phase.

\begin{table}[h!]
\centering
\begin{tabular}{ c | c  c  c } 
 \hline
 \multicolumn{4}{c}{Anode solid phase} \\ [0.5ex] 
 \hline
 Composition & Weight\% & Density [g/cm$^3$] & Volume\% \\
 \hline
 \hline
 Graphite & 94.5 & 2.26 & 92.6 \\
 \hline
 CMC & 2.25 & 1.6 & 3.1 \\
 \hline
 SBR & 2.25 & 1.52 & 3.3 \\
 \hline
 Carbon black & 1 & 2.1 & 1.1 \\
 \hline
 \multicolumn{4}{c}{Cathode solid phase} \\ [0.5ex]
 \hline
 Composition & Weight\% & Density [g/cm$^3$] & Volume\% \\
 \hline
\hline
 LiFePO$_4$ & 93.5 & 3.5 & 89.0 \\
 \hline
 PVDF & 2.5 & 1.78 & 4.7 \\
 \hline
 Carbon black & 4 & 2.1 & 6.3 \\ [1ex] 
 \hline
\end{tabular}
\caption{Electrode solid phase volume fraction. \tablefootnote{https://www.mtixtl.com/Li-ionBatteryChemicalPowdersBindersandElectrodesSheet.aspx}}
\label{tab:binder}
\end{table}

By referring to Tables \ref{tab:binder} and \ref{tab:vol} and assuming a porosity for each porous material, the total thickness of each layer type was obtained, as is listed in Table \ref{tab:tof}. Then, given the wave velocity in each material, the ToF of ultrasound transmitted through the cell can be obtained.

\begin{table}[h!]
\centering
\begin{tabular}{ c | c  c  c  c  c  c} 
 \hline
 Composition & LiFePO$_4$ & Electrolyte & Graphite & Al & Cu & PP \\ [0.5ex] 
 \hline
 \hline
 Weight\% \tablefootnote{Material Safety Data Sheet (MSDS) of Li-ion battery LF50K, EVE POWER Co., Ltd. https://www.evebattery.com/prismatic-lfp-cell} & 31 & 22 & 17 & 18 & 10 & 2 \\
 \hline
 Density [g/cm$^3$] & 3.5 & 1.27 \cite{gold2017probing} & 2.26 & 2.7 & 8.96 & 0.9 \\
 \hline
 Volume\% & 21.1 & 38.2 & 18.6 & 14.7 & 2.5 & 4.9 \\ [1ex] 
 \hline
\end{tabular}
\caption{Composition volume fraction of the prismatic cell.}
\label{tab:vol}
\end{table}

\begin{table}[h!]
\centering
\begin{tabular}{c | c  c  c  c} 
 \hline
 Layer & \makecell[c]{Porosity \\ (Assumed)} & \makecell[c]{Total \\ thickness [mm]} & $c$ [m/s] & ToF [us] \\ [0.5ex] 
 \hline
 \hline
 \makecell[c]{Anode \\ (Graphite\&Electrolyte)} &  0.37 & 9.21 & \makecell[c]{1154.8 \\ (FEM)}  & 7.97 \\
 \hline
 \makecell[c]{Cathode \\ (LiFePO$_4$\&Electrolyte)} & 0.4  & 10.91 & \makecell[c]{1145.4 \\ (FEM)}  & 9.53 \\
 \hline
 \makecell[c]{Separator \\ (PP\&Electrolyte)} & 0.4  & 2.54  & \makecell[c]{1353.7 \\ (FEM)} & 1.88 \\
 \hline
 Cu &  & 0.77 & 4600 & 0.17 \\ 
 \hline
 Al &  & 5.87 & 6320 & 0.93 \\
 \hline
 Total & & 29.3 & & 20.47 \\ [1ex] 
 \hline
\end{tabular}
\caption{Total thicknesses of different layers of the prismatic cell and the calculated ToFs with the FEM model.}
\label{tab:tof}
\end{table}

\begin{figure}
    \centering
    \includegraphics{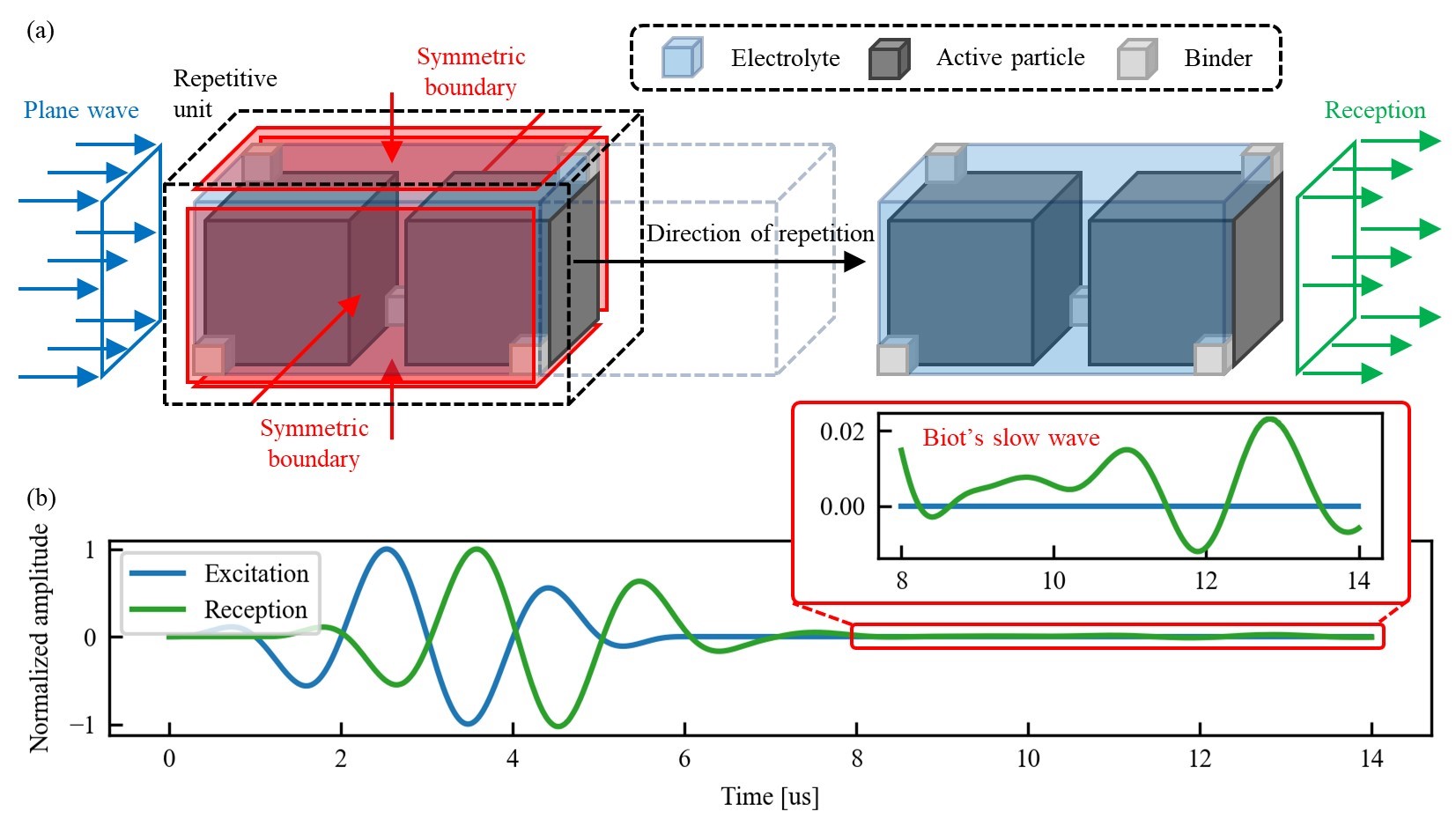}
    \caption{(a) Schematic of the geometry of the 3-D FEM model for electrodes and (b) an example received time-domain signal given a 1.2 mm propagation distance.}
    \label{fig:fem}
\end{figure}

The wave velocities in electrodes were obtained with 3-D FEM. The modelling microstructure of an electrode is shown in Figure \ref{fig:fem}(a). The active material particles are modelled as arrayed cubes with a representative size of 10 $\mu$m. The binder and the conductive carbon black are modelled as a single material, in the form of cuboids that connect the particles, considering that the conductive carbon black is in the form of nanoparticles and mixed with binder material in the micro-scale. The remaining regions are electrolytes. A 3-D model is essential since, in practical scenarios, the fluid phase is continuous. The modelling microstructure is a simplified version of reality. However, simulations with different microstructures confirm that the variation in the microstructure has little effect on the wave velocity (see \ref{sec:micro:appendix}).

On the left end of the model, a Hanning-windowed 3-cycle sinusoidal transient load was applied, as shown in Figure \ref{fig:fem}(b). The central frequency of the load is 500 kHz, which is close to the central frequency of measured signals in the experiment (see Figure \ref{fig:result}(a)). The waves propagated through the model were recorded on the right end. Considering that the electrodes have large dimensions perpendicular to the wave propagation direction, the plane strain condition should be met in these dimensions. Therefore, we only model a unit volume in these dimensions, and symmetric boundary conditions are applied. On the other hand, the dimension of the direction of propagation should be large enough to ensure the separation between the first arrived longitudinal wave and another slow wave predicted by Biot theory. The slow wave has a high attenuation rate in practice, and we only care about the fast wave in this study. Then, the ToF of the longitudinal wave through the model can be estimated with a received longitudinal wave's waveform that is not polluted by the slow wave.

In the FEM model, the acoustic behaviour in a solid is described by
\begin{equation}
    \rho_s \ddot{\mathbf{u}} + \nabla \mathbf{\sigma} = 0,
\end{equation}
where $\rho_s$ is the density of the solid, $\mathbf{u}$ is the displacement vector, and $\mathbf{\sigma}$ is the Cauchy stress tensor.

For the acoustic behaviour in the fluid phase, as we only study the wave velocity, the losses due to viscosity and thermal conduction will be neglected. The governing equation for the fluid phase is
\begin{equation}
    \frac{1}{\rho_f c_f ^ 2} \ddot{p} + \nabla \cdot \left( -\frac{1}{\rho_f} \nabla p \right) = 0,
\end{equation}
where $p$ is the sound pressure; $\rho_f$ and $c_f$ are the density and sound speed of the fluid.

At the fluid-solid interfaces, the sound pressure and acceleration are continuous. The friction between the fluid and the solids due to fluid viscosity is neglected as it has little influence on the wave velocity \cite{santos2016waves}. In this study, the coupling is realized by the COMSOL Acoustic-Solid Interaction, Transient Interface, which is described by
\begin{eqnarray}
-\mathbf{n}\cdot\left( -\frac{1}{\rho_f} \nabla p \right)&=&- \mathbf{n} \cdot \ddot{\mathbf{u}} \\
\mathbf{F}_A&=&p\mathbf{n}
\end{eqnarray}
where $\mathbf{n}$ is the surface normal of a fluid-solid interface, and $\mathbf{F}_A$ is the load force per unit area experienced by the solids.

The aforementioned equations were solved by 3-D FEM with the material properties as listed in Table \ref{tab:mech}. The bulk and shear modulus of the binder/carbon composite materials are estimated by referring to the measured elastic modulus in the literature \cite{grillet2016conductivity, hu2019effect} and assuming a Poisson's ratio of 0.35. This value was chosen because plastics usually have a Poisson's ratio of 0.4, while the one of carbon black is 0.3.

\begin{table}[h!]
\centering
\begin{tabular}{ c | c  c  c } 
 \hline
 Composition & Bulk modulus [GPa] & Shear modulus [GPa] \\ [0.5ex]
 \hline
 \hline
 Graphite \cite{qi2010threefold} & 27 & 12.5 \\
 \hline
 \makecell[c]{Anode binder \cite{hu2019effect} \\ (CMC\&SBR\&Carbon black)} & 0.56 & 0.19 \\
 \hline
 LiFePO$_4$ \cite{maxisch2006elastic} & 95 & 45 \\
 \hline
 \makecell[c]{Cathode binder \cite{grillet2016conductivity} \\ (PVDF\&Carbon black)} & 1.78  & 0.59 \\
 \hline
 PP & 1.38 & 0.92 \\
 \hline
 Electrolyte \cite{gor2014model} & 1 &  \\ [1ex] 
 \hline
\end{tabular}
\caption{Composition mechanical properties of the prismatic cell}
\label{tab:mech}
\end{table}

Figure \ref{fig:fem}(b) shows an example time-domain signal received on the right end, in which the separated longitudinal wave and slow wave can be observed.

The FEM-calculated ToF through electrodes of the prismatic cell is listed in Table \ref{tab:tof}. The wave velocity in the separator was also calculated by FEM and the solid phase was modelled as a cubic lattice. This FEM-calculated velocity, 1353.7 m/s, was found to be in good agreement with the one based on Biot theory (1345.4 m/s). Further details regarding the application of Biot theory can be found in \ref{sec:biot:appendix}. Overall, the total ToF through the cell, calculated as 20.5 $\mu$s, was found to agree well with the experiment value of 20.1 $\mu$s.

\subsection{The effect of binder degradation}

\begin{figure}
    \centering
    \includegraphics{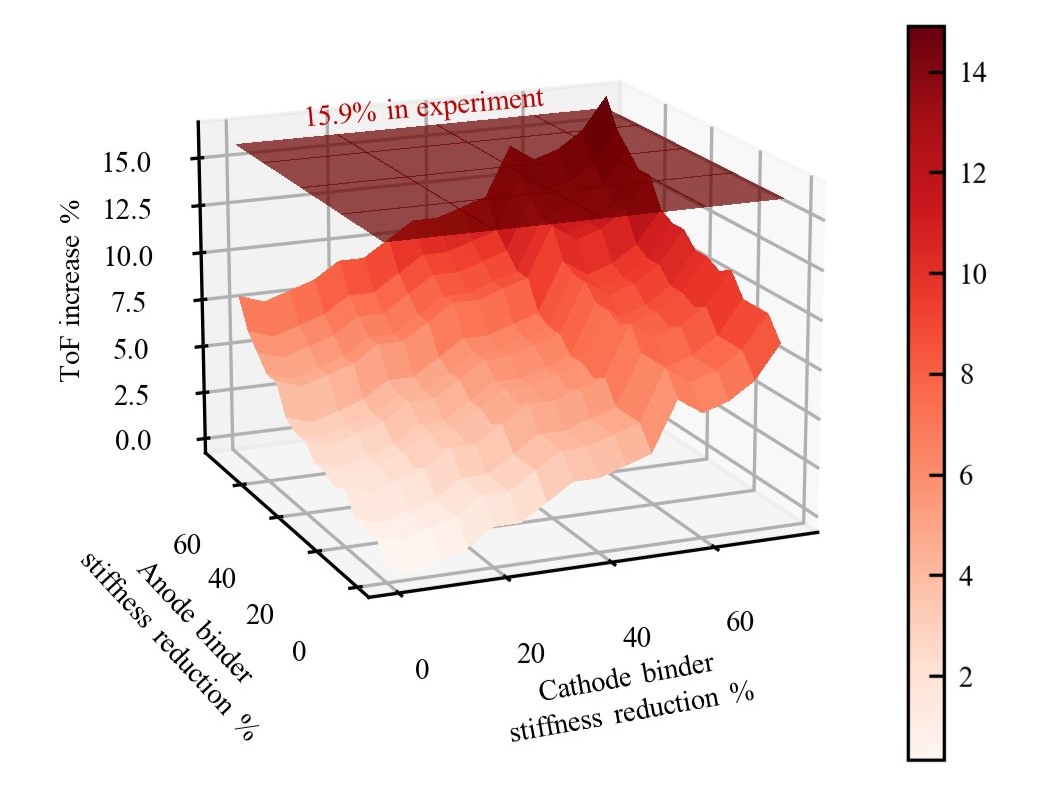}
    \caption{The effect of binder degradation on ultrasound ToF evaluated by FEM.}
    \label{fig:binder}
\end{figure}

The ToF corresponding to a bulk and shear modulus reduction from 0\% to 75\% of the binder/carbon composite material was calculated by the FEM model. As shown in Figure \ref{fig:binder}, when the stiffness reductions for both anode and cathode were up to 75\%, the ToF increase reached 16.5\%. This increase is comparable to the one observed in experiments.

Based on the experiments listed in Table \ref{tab:nano}, the stiffness reduction of the LFP cathode can reach up to 76.1\% under 2 C-rate and 90\% capacity retention. Higher C-rate and number of cycles lead to greater stiffness reduction. In our experiments, we observed a 15.9\% increase in ToF under the conditions of 3 C-rate and 87.6\% capacity retention. Therefore, the electrode stiffness reduction should be comparable to or even larger than 76.1\%. In the FEM model, a 75\% binder stiffness reduction leads to a 74.4\% reduction in anode elastic modulus and a 74.3\% reduction in cathode elastic modulus. Therefore, it is reasonable to conclude that the observed 15.9\% increase in ToF can be attributed to these stiffness reductions.

\begin{table}[h!]
\centering
\begin{tabular}{ c | c  c  c | c  c } 
 \hline
 Sample & \multicolumn{3}{c|}{Samples from \cite{demirocak2015probing}} & \multicolumn{2}{c}{Our sample} \\ [0.5ex] 
 \hline
 \hline
 Electrode & Cathode & Cathode & Cathode & Cathode & Anode \\
 \hline
 \makecell[c]{Ambient \\ temperature [$^\circ$C]} & 55 & 55 & 55 & 35 & 35 \\
 \hline
 C-rate & 1 & 1 & \makecell[c]{1 (charge) \\ 2 (discharge)} & 3 & 3 \\
 \hline
 \makecell[c]{Capacity \\ retention\%} & 90 & 80 & 90 & 87 & 87  \\
 \hline
 \#Cycles & 459 & 517 & 171 & 500 & 500 \\
 \hline
 \makecell[c]{Elastic modulus \\ reduction\%} & \makecell[c]{23.4 \\ (Measured)} & \makecell[c]{41.3 \\ (Measured)} & \makecell[c]{76.1 \\ (Measured)} & \makecell[c]{74.3 \\ (Assumed)} & \makecell[c]{74.4 \\ (Assumed)} \\ [1ex] 
 \hline
\end{tabular}
\caption{Elastic modulus reduction of the dried electrodes of LFP prismatic cells under different cycling conditions. }

\label{tab:nano}
\end{table}

\subsection{The effect of loss of active lithium}

The loss of active lithium in batteries can occur through three different mechanisms: SEI growth, lithium plating, and LAM. To quantify the effects of these fading mechanisms individually, we assume a 10\% lithium loss, which is attributed solely to one of these mechanisms. Moreover, the ageing effects are quantified at a fully-discharged state.

To investigate the effect of SEI growth or lithium plating, we assume that the SEI or lithium is uniformly deposited on the surface of graphite particles, forming a thin layer with no change in the thickness of the anode layer. By referring to the composition volumes listed in Table \ref{tab:vol} and the molar volumes of graphite, SEI \cite{borodin2006molecular,heiskanen2019generation}, and lithium, we obtained a graphite particle size increased by 4.2\% and 0.6\% due to SEI growth and lithium plating, respectively. This calculation agrees with the experiments that reveal the SEI thickness is usually in the nano-scale \cite{park2012principles,you2023situ}. On the other hand, the loss of active lithium reduces lithium intercalation in cathode active particles, leading to a 2\% reduction in bulk modulus and a 0.5\% reduction in volume. As the anode layer thickness remains constant, we assume that the cathode layer experiences a free thickness shrinkage due to the reduction in the volume of active particles. Then, the thickness change of the cathode layer was obtained based on the porosity listed in Table \ref{tab:tof}. The space created by the cathode shrinkage can be filled with electrolytes. Overall, either SEI growth or lithium plating causes only negligible mechanical property changes.

For LAM, we consider the detachment of active materials from the binder/carbon matrix. In this case, the lithium in the detached active material particles would be blocked as the particles become non-conductive. Here, we made a quantification by assuming the lost 10\% lithium is blocked in the detached graphite particles. At the fully-discharged state, the stiffness of graphite particles remains constant, but the particle size increases by 1.5\% compared to the case without lithium. Regarding the layer thicknesses, as the graphite expansion is larger than the cathode particles' shrinkage, the anode would not experience a free expansion. Instead, it would be pressed by the separator and the cathode layer, and the binders in these layers would suffer compression. Given the binder volumes and the bulk modulus of the binders and the separator, the equilibrium layer thickness variations are -0.192\% and +0.233\% for the cathode and anode, respectively. In summary, the LAM also leads to negligible mechanical property changes.

Overall, the factors related to the re-distribution of lithium should have limited contribution to the change of ultrasound ToF, as summarized in Table \ref{tab:effect2}. This finding agrees with our experimental results, showing little ToF variation over SOC and with capacity regeneration.

\begin{table}[h!]
\centering
\begin{tabular}{c c | c  c c} 
 \hline
 \multicolumn{2}{c|}{ } & SEI growth & Lithium plating & LAM \\ [0.5ex] 
 \hline
 \hline
 Cathode & Bulk modulus \cite{maxisch2006elastic} &  -2\%  & -2\% & -2\% \\
 & Particle size \cite{maxisch2006elastic} & -0.167\%  & -0.167\% & -0.167\% \\
 & Layer thickness & -0.135\%  & -0.135\% & -0.192\% \\
 \hline
 Anode & Particle size & +4.2\% & +1.1\% & +1.5\% \cite{qi2010threefold} \\
 & Layer thickness & 0 & 0 & +0.233\% \\ [1ex] 
 \hline
\end{tabular}
\caption{Mechanical property change due to 10\% loss of active lithium.}
\label{tab:effect2}
\end{table}

\subsection{The effect of gas generation or unwetting}

The SEI growth and electrolyte consumption would introduce dilute gas bubbles in the electrolyte or gas regions in the pores \cite{deng2020ultrasonic}. For dilute gas bubbles, extended Biot theories have been developed by modifying the liquid parameters for the gas-liquid mixture \cite{corapcioglu1991wave,smeulders1997wave}. In the ultrasonic frequency range (larger than 100 kHz), gas bubbles affect the sound speed only if their size is smaller than 2.5 $\mu$m (see \ref{sec:bubble}). However, the microbubbles lose gas and disappear quickly \cite{klibanov2006microbubble,martin2013current}. For the gas region in pores, the cell in our experiment has extra electrolytes and the whole electrodes are fully soaked in electrolytes, and the gas region should be replenished by the extra electrolytes. Therefore, even if the gas generation affects the ultrasound speed, the effect should be short-term and recoverable. This means that long-term measurements can be used to reveal the effect of gas generation.

We performed ultrasonic testing on the aged cell after long-term storage and compared the ultrasonic signal with the one recorded right after the cycling under room temperature. The ToF of transmitted ultrasound for the two cases is 20.9 $\mu$s and 20.5 $\mu$, respectively, which has a decrease of 2\%. It demonstrates gas generation or unwetting should not be a major factor responsible for the ToF variation.

\subsection{Discussion}
In summary, binder degradation can induce a comparable ToF increase with the experiment while the loss of active lithium only has negligible contributions, and the gas in the electrodes can not have a significant and unrecoverable influence on the ToF.

The binder degradation that causes the ToF increase does not have a direct correlation with the capacity fading. Instead, it should reveal the cycling history, such as the C-rate and number of cycles. It implies that the ToF may not have a stable correlation with capacity retention.

We performed extra cycling on the cell and found that, although the capacity kept fading at the same rate, ToF remained constant at later cycles, which results in the increase of PCC from -0.981 to -0.917. This is because binder degradation mainly occurs in the early cycles, whereas the loss of active lithium remains at a constant rate throughout all cycles.

Therefore, if we use ToF as the feature to predict capacity retention, it may fail at late cycles. Instead, the ToF feature should be useful for predicting cycling history because the C-rate and number of cycles have a direct correlation with binder degradation.

As for the generality of the correlation between ToF and SoH, despite the various cell types and operating conditions such as temperature, C-rate, and depth of charge/discharge, the binder that connects active materials should always experience stiffness reduction due to the expansion/shrinkage of active materials during lithium intercalation/deintercalation. For example, both Davies \textit{et al.} \cite{davies2017state} and Wu \textit{et al.} \cite{wu2019ultrasonic} observed the increase of ultrasound ToF along with the number of cycles of LCO pouch cells. However, the rate of stiffness reduction should be dependent on the cell types and operating conditions. The accuracy of SoH prediction with ToF would be affected subsequently.

\section{Conclusion}
\label{sec:con}

In this work, there are mainly two original contributions. Firstly, we confirmed the correlation between ultrasound speed and SoH of an LFP prismatic cell with an ultrasonic battery health monitoring experiment. During the 3 C-rate cycling of the cell, both the capacity retention and ultrasound speed kept decreasing, At the end of 500 cycles, the capacity decreased by 12.4\% while the ultrasonic transmission ToF increased by 15.9\%. Secondly, we explained the correlation by constructing a mathematical model for calculating the transmission ToF through the cell, in which the wave velocity in electrodes was obtained by simulating the fluid-solid interaction on a micro-scale and wave propagation in the fluid-solid system with 3-D FEM. The model also agrees with Biot theory in terms of the velocity in the separator. With the wave propagation model, we found that the ToF is largely affected by the binder stiffness reduction due to cycling, compared with the SEI growth, lithium plating, LAM, and gas bubbles/regions. It implies that the ultrasound ToF feature is suitable for cycling history prediction rather than predicting capacity retention. An additional model that correlates cycling history with capacity retention is needed to predict capacity retention with ToF.

Further experimental confirmations are required in the future to make the conclusion more credible. Firstly, mechanical testing on the pristine and degraded electrodes is needed to verify the assumption made in this paper, that is, the binder stiffness suffers a 75\% reduction. Also, we may perform ultrasonic monitoring on a cell during its SEI formation cycles to verify the effect of SEI and gas generations because a large amount of SEI and gas will be generated. This experiment will be conducted on lab-made cells because the commercialized cells have already experienced formation cycles.

\section*{Acknowledgement}
This research is supported by MOE AcRF Tier 1, RG69/22 and A*STAR (project no. C210812038).

\appendix

\section{Experimental results on another cell}
\label{sec:exp2:appendix}

\begin{figure}
    \centering
    \includegraphics{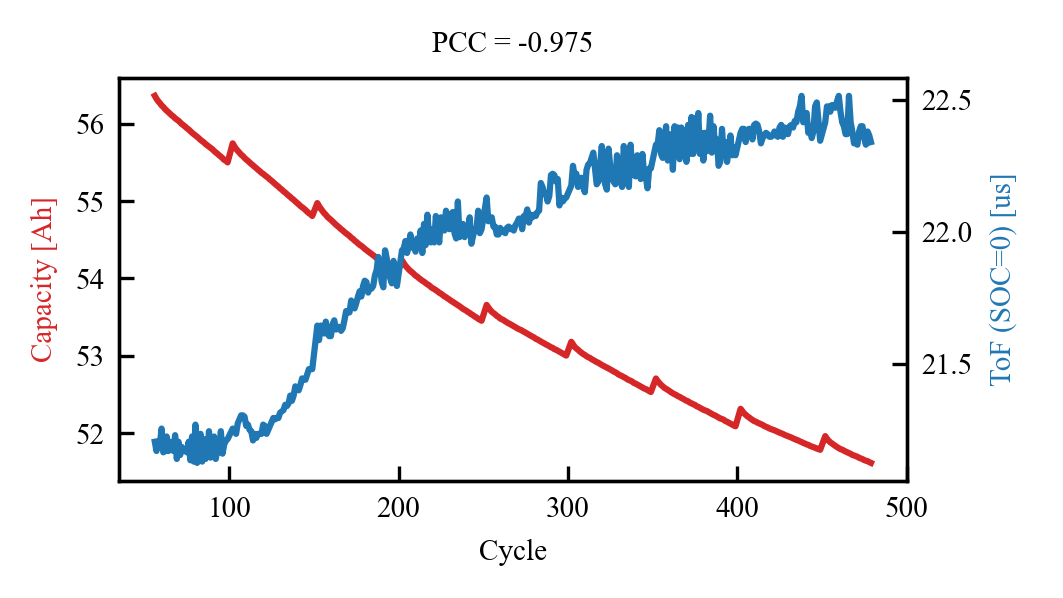}
    \caption{Capacity retention and ToF over cycles for another cell.}
    \label{fig:result2}
\end{figure}

Another cell with the same type was also cycled at a 3 C-rate. It also experienced resting stages every 50 cycles. The capacity retention and ToF at SOC=0 over cycles are shown in Figure \ref{fig:result2}. The capacity experienced an 8.5\% decrease from 56.4 Ah to 51.6 Ah while the ToF experienced a 5.7\% increase from 21.2 $\mu$s to 22.4 $\mu$s. This result is similar to the one shown in Figure \ref{fig:result}.

\section{The effect of microstructure in the FEM model}
\label{sec:micro:appendix}

The microstructure we adopted is with a 10 $\mu$m particle size and each particle connects with six adjacent particles, i.e., a simple cubic structure. Here, we varied either the particle size or structure to investigate the effect of microstructure on the ultrasound speed.

\begin{figure}
    \centering
    \includegraphics{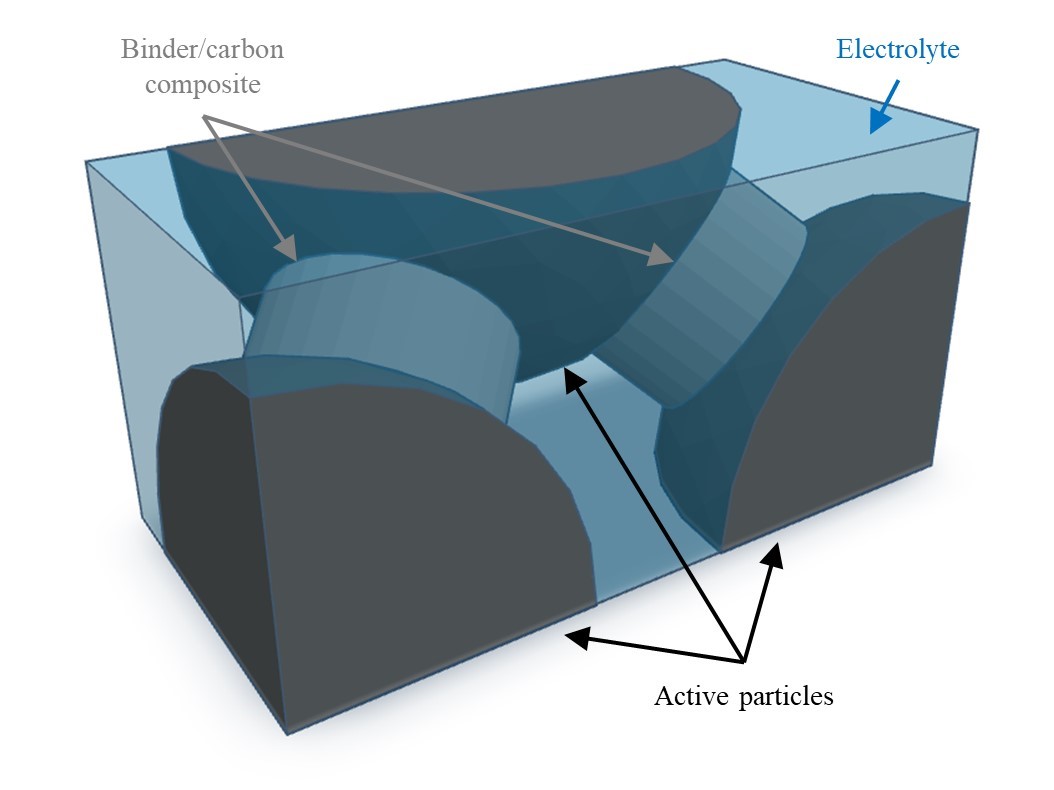}
    \caption{The repetitive unit of the electrode FEM model with a body-centred structure.}
    \label{fig:structure}
\end{figure}

At first, a body-centred structure was evaluated for both electrodes. As shown in Figure \ref{fig:structure}, the particles are with a sphere geometry and each particle connects with eight adjacent particles. The number of connections is larger than the one with a simple cubic structure, which leads to a larger stiffness of the electrodes. The larger stiffness can be reflected in the ToF calculation results listed in Table \ref{tab:structure}, which show that the ultrasound speeds in both electrodes are slightly larger than the ones with a simple cubic structure.

Moreover, as LFP particles typically have a representative size of 3 $\mu$m, we modelled the cathode with this particle size and found that ToF had little change. Overall, the microstructure should have little effect on the ultrasound ToF.

\begin{table}[h!]
\centering
\begin{tabular}{c | c  c  c  c} 
 \hline
 Microstructure & \makecell[c]{Fresh \\ anode} & \makecell[c]{Aged \\ anode} & \makecell[c]{Fresh \\ cathode} & \makecell[c]{Aged \\ cathode} \\ [0.5ex] 
 \hline
 \hline
 \textbf{\makecell[c]{10 um particle size \& \\ Simple cubic structure}} & \textbf{1154.8} & \textbf{960.3} & \textbf{1145.5} & \textbf{966.0} \\
 \hline
 \makecell[c]{10 um particle size \& \\ Body-centered cubic structure} & \makecell[c]{1172.9 \\ (+1.5\%)} & \makecell[c]{991.2 \\ (+3.2\%)} & \makecell[c]{1185.5 \\ (+3.5\%)} & \makecell[c]{951.3 \\ (-1.5\%)} \\
 \hline
 \makecell[c]{3 um particle size \& \\ Simple cubic structure} &  &  & \makecell[c]{1148.7 \\ (+0.3\%)} & \makecell[c]{985.5 \\ (+2\%)} \\ [1ex] 
 \hline
\end{tabular}
\caption{Effect of microstructure on sound speed [m/s]; 'Fresh' and 'Aged' represent without and with 75\% binder stiffness reduction, respectively.}
\label{tab:structure}
\end{table}

\section{Biot theory}
\label{sec:biot:appendix}

The elementary volume for study in Biot theory is a macroscopic cubic and the displacements are the average of microscopic values. The material of the frame is regarded as an isotropic linear elastic solid on the macro-scale. The stress-train relations are

\begin{eqnarray}
\sigma_{ij}^{(m)}&=&2G_md_{ij}^{(m)}+(Ke_m+Ce_f)\delta_{ij}, \\
\sigma_f&=&Ce_m + Re_f,
\end{eqnarray}
where
\begin{eqnarray}
    K&=&\frac{(1-\phi)\left(1-\phi-K_m/K_s\right)K_s+\phi K_s K_m /K_f}{1-\phi-K_m/K_s+\phi K_s/K_f}, \\
    C&=&\frac{\phi\left(1-\phi-K_m/K_s\right)K_s}{1-\phi-K_m/K_s+\phi K_s/K_f}, \\
    R&=&\frac{\phi^2K_s}{1-\phi-K_m/K_s+\phi K_s/K_f},
\end{eqnarray}
$\phi$ is the porosity, $\delta_{ij}$ is the Kronecker delta, $\sigma_{ij}^{(m)}$ is the stress applied on the solid matrix; $d_{ij}^{(m)}$, $e_m$, $K_m$, and $G_m$ are the distortion, dilation, bulk modulus, and shear modulus of the solid matrix, respectively; $K_s$ is the bulk modulus of the solid material; $\sigma_f$, $e_f$, and $K_f$ are the fluid pressure, dilation, and bulk modulus, respectively.

For the separator, $K_m$ and $G_m$ are approximated by \cite{huang2022quantitative}

\begin{equation}
    K_m = 4G_sK_s(1-\phi)/(4G_s+3\phi K_s)
\end{equation}
and
\begin{equation}
    G_m = G_s(8G_s+9K_s)(1-\phi)/(8G_s+9K_s+6(2G_s+K_s)\phi),
\end{equation}
where $G_s$ is the shear modulus of the solid material.

Then, the equation of motion for longitudinal waves in the lossless case is

\begin{equation} \label{eq:motion}
    \partial_i\partial_j\sigma_{ij}^{(m)}=\rho_{11}\ddot{e}_m+\rho_{12}\ddot{e}_f,
\end{equation}
where

\begin{eqnarray}
    \rho_{12}&=&-\phi \rho_f(\tau-1), \\
    \rho_{11}&=&(1-\phi)\rho_s - \rho_{12}, \\
    \rho_{22}&=&\phi\rho_f -\rho_{12},
\end{eqnarray}
and $\tau$ is tortuosity to account for the microscopic inertial interactions between solid and fluid.

Assume plane wave solution and $e_{m0}$ and $e_{f0}$ are the wave amplitudes, Equation (\ref{eq:motion}) becomes

\begin{eqnarray}
    \mathbf{B}\mathbf{e}&=&c_p^2\mathbf{D}\mathbf{e} \\
    \Longrightarrow
    \begin{bmatrix}
    K+4G_m/3 & C \\
    C & R
    \end{bmatrix}
    \begin{Bmatrix}
    e_{m0} \\
    e_{f0}
    \end{Bmatrix}
    &=&c_p^2
    \begin{bmatrix}
        \rho_{11} & \rho_{12} \\
        \rho_{12} & \rho_{22}
    \end{bmatrix}
     \begin{Bmatrix}
    e_{m0} \\
    e_{f0}
    \end{Bmatrix}.
\end{eqnarray}
where $c_p$ is the phase velocity. The above equation has non-trivial solutions of $\mathbf{e}$ if

\begin{equation}
    \text{det}\left(\mathbf{D}^{-1}\mathbf{B}-c_p^2\mathbf{I}\right) = 0.
\end{equation}
The solution of this second-order equation in $c_p^2$ has two roots, corresponding to two longitudinal waves; the faster one is a wave for which the solid and the fluid move in phase, and another in which they are in counter-phase, a.k.a. Biot's slow wave.

\section{The effect of gas bubbles on wave velocity}
\label{sec:bubble}

The phase velocity of acoustic waves in a bubbly liquid can be described by \cite{commander1989linear}
\begin{equation}
    \frac{c^2}{c_m^2} = 1 + \frac{4\pi c^2nr}{\omega_0^2-\omega^2+2ib\omega},
\end{equation}
where $c$ is the velocity in the liquid, $c_m$ is the velocity in the bubbly liquid, $n$ is the number of bubbles per unit volume, $r$ is the radius of the bubbles, $\omega_0$ is the resonant angular frequency of the bubbles, $\omega$ is the angular frequency of the waves, and $b$ is the damping constant of the bubbles.

The $\omega_0$ and $b$ are given by the following equations:
\begin{equation}
    \omega_0^2=\frac{p_0}{\rho r^2}\left({\rm Re}\Phi-\frac{2\sigma}{rp_0}\right)
\end{equation}
and
\begin{equation}
    b = \frac{2\mu}{\rho r^2} + \frac{p_0}{2\rho r^2} {\rm Im} \Phi +\frac{\omega^2r}{2c},
\end{equation}
where $p_0$ is the ambient pressure, $\rho$ is the density of the liquid, $\sigma$ is the surface tension of the liquid, $\mu$ is the viscosity, and $\Phi$ is described by
\begin{equation}
    \Phi = \frac{3\gamma}{1-3(\gamma-1)i\chi [(i/\chi)^{1/2}\coth(i/\chi)^{1/2}-1]},
\end{equation}
where $\chi=D/\omega r^2$, $D$ is the gas thermal diffusivity, and $\gamma$ is the ratio of specific heats.

\begin{figure}[h]
    \centering
    \includegraphics{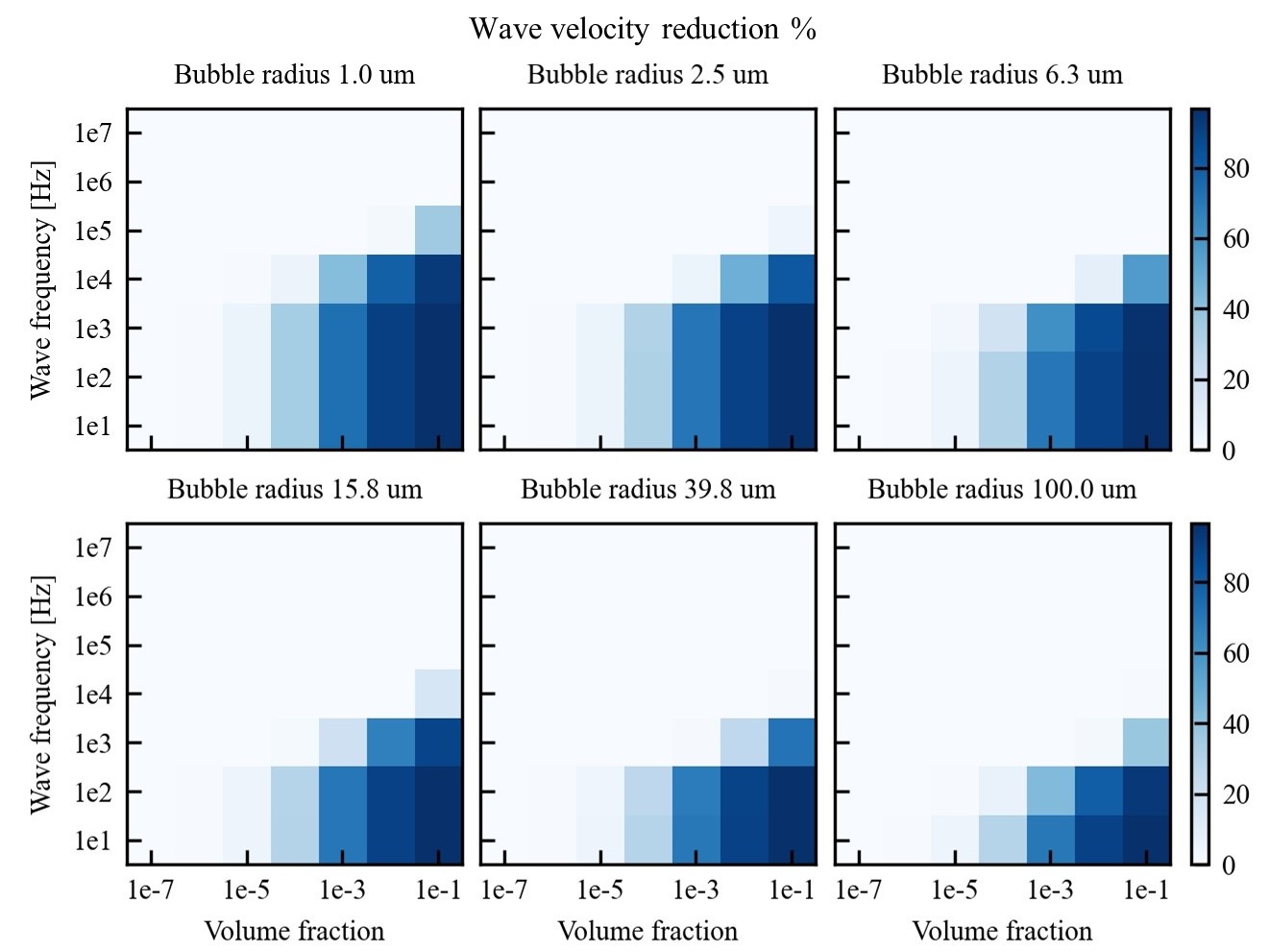}
    \caption{Phase velocity reduction of acoustic waves due to the existence of dilute bubbles in liquid versus bubble radius, frequency, and gas volume fraction.}
    \label{fig:bubble}
\end{figure}

The phase velocity calculated by these equations is shown in Figure \ref{fig:bubble}. The effect of bubbles on the phase velocity is a reducing effect.

From the results, we can conclude that smaller bubbles, larger volume fraction, and lower frequency lead to a larger decrease in the phase velocities. In the ultrasonic frequency range (larger than 100 kHz), there is no velocity-decrease effect unless the bubble radius is smaller than 2.5 um.

 \bibliographystyle{elsarticle-num} 
 \bibliography{refs}





\end{document}